\documentclass[final,authoryear,5p]{elsarticle}

\usepackage{epsfig}
\usepackage{amsmath}
\usepackage{amssymb}
\usepackage{graphicx}			
\usepackage{dcolumn}			
\usepackage{bm}					
\usepackage{esdiff}				
\usepackage{hyperref}			
\usepackage[mathlines]{lineno}	
\usepackage{esdiff}
\usepackage{xcolor}

\journal{Advances in Space Research}

\begin{document}

\begin{frontmatter}

\title{Modeling approaches for precise relativistic orbits: \\ Analytical, Lie-series, and pN approximation}
\author[ZARM]{Dennis Philipp\corref{cor}}
\cortext[cor]{Corresponding author}
\ead{dennis.philipp@zarm.uni-bremen.de}

\author[ZARM]{Florian Woeske}
\author[IfE]{Liliane Biskupek}
\author[ZARM]{Eva Hackmann}
\author[IfE]{Enrico Mai}
\author[ZARM]{Meike List}
\author[ZARM]{Claus L{\"a}mmerzahl}
\author[ZARM]{Benny Rievers}

\address[ZARM]{ZARM, University of Bremen, 28359 Bremen, Germany}
\address[IfE]{IfE, Leibniz University, 30167 Hannover, Germany}

\begin{abstract}
Accurate orbit modeling plays a key role in contemporary and future space missions such as GRACE and its successor GRACE-FO, GNSS, and altimetry missions. To fully exploit the technological capabilities and correctly interpret measurements, relativistic orbital effects need to be taken into account. 

Within the theory of General Relativity, equations of motion for freely falling test objects, such as satellites orbiting the Earth, are given by the geodesic equation. We analyze and compare different solution methods in a spherically symmetric background, i.e.\ for the Schwarzschild spacetime, as a test bed. We investigate satellite orbits and use direct numerical orbit integration as well as the semi-analytical Lie-series approach. The results are compared to the exact analytical reference solution in terms of elliptic functions. For a set of exemplary orbits, we determine the respective accuracy of the different methods.

Within the post-Newtonian approximation of General Relativity, modified orbital equations are obtained by adding relativistic corrections to the Newtonian equations of motion. We analyze the accuracy of this approximation with respect to the general relativistic setting. Therefore, we solve the post-Newtonian equation of motion using the eXtended High Performance Satellite dynamics Simulator. For corresponding initial conditions, we compare orbits in the Schwarzschild spacetime to those in its post-Newtonian approximation. Moreover, we compare the magnitude of relativistic contributions to several typical perturbations of satellite orbits due to, e.g., solar radiation pressure, Earth's albedo, and atmospheric drag. This comparison is done for our test scenarios and for a real GRACE orbit to highlight the importance of relativistic effects in geodetic space missions. For the considered orbits, first-order relativistic contributions give accelerations of about $20\,$nm/s$^2$ and are dominant in the radial direction.
\end{abstract}

\begin{keyword}
Relativistic geodesy; post-Newtonian theory; Relativistic effects; Satellite orbits; Orbit propagation \\
\textit{PACS:} 91.10.-v; 91.10.By; 91.10.Fc; 91.10.Sp; 04.25.Nx
\end{keyword}

\end{frontmatter}
\parindent=0.5 cm


\section{\label{Sec_intro} Introduction}

Contemporary and future high precision geodesy and gravimetry space missions require a precise modeling of satellite orbits. Missions such as GRACE-FO, the successor of the long-lasting gravity field recovery mission GRACE \citep{Tapley:2004}, aim at nanometer accuracy in the change of the spatial distance between two spacecraft  \citep{Sheard:2012,Loomis:2012,Flechtner:2016}. At this level of accuracy, relativistic effects need to be taken into account. Therefore, precise orbit modeling and orbit propagation tools, incorporating relativistic equations of motion, are needed to consistently interpret measurements at the best possible level of accuracy. These tools usually use a numerical integration procedure and our main goal is to quantify the accuracy of this approach in two different ways outlined below.

For satellite missions and measurements, that impose a high position or orbit accuracy, accurate modeling of all physical effects acting on a satellite is essential to fully exploit measurement data. Accurate and validated force models are the basis for mission design, analysis, as well as precise orbit determination (POD) \citep{Wu:1991,Jaeggi:2006}. POD techniques give the most accurate orbit estimates from all kind of measurement data. Nowadays, measurement methods and sensors reached an unprecedented precision such that relativistic effects must be considered, having nearly the same order of magnitude as solar radiation pressure (SRP) for conventional low Earth orbit (LEO) satellites.
Exemplary missions for which highest position accuracy is required are altimetry, GNSS, geodesy, and fundamental science. For GNSS satellites, the absolute accuracy is about 5 to 30 cm \citep{Steigenberger:2015}. For the GRACE satellites, where more measurement data are constantly available, an accuracy of less then 5 cm is achieved \citep{Kang:2003,Kang:2006}.

Within the theory of General Relativity (GR), freely falling test bodies move on timelike geodesics and the equation of motion (EOM) is given by the geodesic equation, which involves quantities derived from the spacetime metric, see, e.g., \citep{MTW:1973}. For a certain class of spacetimes, which are exact solutions of Einstein's field equation with a sufficient amount of symmetries, this equation can be solved analytically, see \citep{Forsyth:1920, Morton:1921, Hagihara:1930, Darwin:1959, Darwin:1961, Hackmann:2009}. Here, we choose one of these spacetimes to build trust into numerical and semi-analytical solution methods, which can be used for more complicated situations where we do not have analytical solutions at hand. We check these methods against an analytical reference solution and investigate their respective accuracy for satellite orbits around the Earth. Our approach is a first step to tackle complex but more realistic situations later on.

The post-Newtonian (pN) EOM used in advanced orbit simulation and propagation tools is an approximation of the general relativistic equation. The eXtended High Performance Satellite dynamics Simulator (XHPS) \citep{Woeske:16} is an orbit propagation tool developed at ZARM, University of Bremen. It numerically solves the (Newtonian) EOM and is also capable of simulating the entire space environment as well as detailed satellite properties. In a second step, we therefore include pN corrections into the XHPS and compare its numerical integrator to the direct numerical integration method applied to the geodesic equation in GR, which we checked before against the analytical solution.

The purpose of this work is twofold: I) For the application within relativistic geodesy, we aim at the comparison of different solution methods for relativistic EOM. II) We investigate the accuracy of the first-order pN approximation of GR for satellite orbits. The pN framework also enables us to compare the magnitude of relativistic effects to various non-gravitational perturbations along satellite orbits.

We use a spherically symmetric gravitational field as a test bed. The general relativistic spacetime is then described by the Schwarzschild metric, and its first-order pN approximation involves the Newtonian gravitational potential of a point mass. This approach only includes the dominant relativistic effect on the orbits, which should, however, be sufficient for a first quantification of the accuracy of orbit simulations within the XHPS and similar tools. For the Schwarzschild spacetime, the exact solutions of the geodesic equation are well-known and given, e.g., in terms of the Weierstrass elliptic function \citep{Hagihara:1930}. All necessary notions are introduced in Sec.\ \ref{Sec_Geometry}, and the EOM is introduced in Sec.\ \ref{Sec_EOM}.

To construct orbits in the Schwarzschild spacetime, we use direct numerical integration, the semi-analytical Lie-series approach, and the analytic solution in terms of the Weierstrass elliptic function. For a pre-defined set of test orbits, the analytical solution serves as the reference to test the accuracy of the other methods. The test orbits and solution methods are introduced in Sec.\ \ref{Sec_SolMethods}, and Sec.\ \ref{Sec_Comparison} contains the results.

We solve the pN EOM using the XHPS that now includes relativistic corrections in the orbit propagation model at the first-order pN level. In Sec.\ \ref{Sec_accuracypN}, we access the accuracy of the pN approximation by analyzing the difference between the XHPS results and the reference orbit with corresponding initial conditions in the Schwarzschild spacetime. 
Finally, in Sec.\ \ref{Sec_RelativisticEffects} we use the XHPS and select one test orbit, for which we assume a GRACE-like satellite model, and a real GRACE orbit from 2008-04-15 to calculate the relativistic accelerations along one full orbital revolution. The result is compared to non-gravitational accelerations due to solar radiation pressure, Earth's albedo, thermal radiation pressure, and atmospheric drag.

Note that for the first-order pN approximation of the Schwarzschild spacetime, the modified Keplerian equations of motion can also be solved analytically to test the accuracy of the pN approximation. However, we use the XHPS since it allows to calculate the magnitude of various non-gravitational perturbations due to the space environment and to show that relativistic effects must be taken into account for high-precision space missions, at least at a pN level.

Table \ref{Tab_methods} shows an overview of the different solution methods for the EOM.

\begin{table}
\centering
	\begin{tabular}{l|c c}
	\hline \hline	
 	\hfil & Schwarzschild \hfil & pN approximation \\
  	\hline
  	EOM & geodesic & pN modified Kepler Eq.\ \\
  	\hline \\
  	method & & \\ 
  	\hline 
  	analytic & reference & n.a. \\
  	Lie-series & test & n.a. \\
  	numerical & test & reference \\
  	XHPS & n.a. & test \\
  	\hline \hline
	\end{tabular}
	\caption{\label{Tab_methods}To construct orbits in the Schwarzschild spacetime, we use the analytical solution as a reference and check the accuracy of a) the semi-analytical Lie-series method and b) the direct numerical integration. To investigate the accuracy of the pN approximation, we solve the modified Keplerian EOM using the XHPS and check the result against a direct numerical integration of the geodesic equation in the Schwarzschild spacetime for corresponding initial conditions and suitable coordinates. In the table, n.a. means that we do not consider the respective points here.}
\end{table}


\section{\label{Sec_Geometry} Geometry and notation}

\subsection{\label{SubSec_Geometry:Schwarzschild} General relativistic spacetime}

Within the theory of GR, the curved spacetime geometry is described by a metric $g$. In a given coordinate system, the metric components are denoted as $g_{\mu\nu}$, where we use greek indices as spacetime indices taking values $0,1,2,3$. The metric itself is to be found as a solution of Einstein's field equation \footnote{Since we aim at applications within relativistic geodesy and for satellite orbits around the Earth, we neglect the influence of a (possible) cosmological constant.}
\begin{align}
\label{Eq_EinsteinFieldEq}
	R_{\mu\nu} - \dfrac{1}{2} R\, g_{\mu\nu} = \dfrac{8\pi G}{c^4} T_{\mu\nu} \, .
\end{align}
Here, $R_{\mu\nu}$ is the Ricci tensor and $R$ is the Ricci scalar. Both are constructed from the spacetime metric, whereas the source term is given by the energy-momentum tensor $T_{\mu\nu}$. The tensorial equation above is a second-order non-linear partial differential equation for the metric components. Newton's gravitational constant $G$ and the speed of light $c$ enter as dimensional factors of proportionality. 

Outside a given mass (energy) distribution, one has to solve the vacuum field equation $R_{\mu\nu} = 0$. The Schwarzschild spacetime is the most famous vacuum solution of Einstein's field equation. It describes the spacetime geometry outside a spherically symmetric mass distribution. This spacetime possesses a monopole moment only, which gives the total mass of the gravitating source. Birkhoff's theorem states that the Schwarzschild spacetime is the unique solution with these properties and spherical symmetry implies that the spacetime must be static. 

Using spherical coordinates $(x^0,r,\vartheta,\varphi)$ and the metric signature convention $(-,+,+,+)$, the Schwarzschild metric reads
\begin{multline}
	\label{Eq_Schwarzschild_metric}
	\mathrm{d}s^2 = - A(r) (\mathrm{d}x^0)^2 + A(r)^{-1} \mathrm{d}r^2 \\
	+ r^2 \mathrm{d}\vartheta^2 + r^2 \sin^2 \vartheta \, \mathrm{d}\varphi^2 \, ,
\end{multline}
where the metric function $A(r)$ is given by
\begin{align}
	\label{Eq_Schwarzschild:A(r)}
	A(r) = 1-2m/r =: 1-r_s/r \, .
\end{align}
The coordinates $x^0$ and $r$ have the dimension of a length, whereas $\vartheta$ and $\varphi$ are the usual angles on the two-sphere $S_2$.
The parameter $m$ is the mass of the gravitating source (in natural units). It is related to the mass $M$ in SI-units by $m=GM/c^2$. The quantity $r_s = 2m$ denotes the Schwarzschild radius (gravitational radius), i.e.\ the radius to which one would have to compress all the mass of the object to form a black hole. For the Earth, the Schwarzschild radius is below one centimeter, $2m_\oplus \approx 0.88\,$cm.

We may explicitly rewrite the Schwarzschild metric \eqref{Eq_Schwarzschild_metric} in SI-units, including $G$ and $c$, and we introduce the coordinate time $t$ by $x^0 =: c\,t$ to obtain
\begin{multline}
	\label{Eq_Schwarzschild_metric_SI}
  	\mathrm{d}s^2 = - \left( 1- \dfrac{2GM}{c^2r} \right) c^2 \mathrm{d}t^2 + \left( 1- \dfrac{2GM}{c^2r} \right)^{-1} \mathrm{d}r^2  \\
		+ r^2 \mathrm{d}\vartheta^2 + r^2 \sin^2 \vartheta \, \mathrm{d}\varphi^2 \, .
\end{multline}
The time coordinate $t$ has the dimension of a time measured in seconds. It will become important in the following sections as a parameter along timelike geodesics of the spacetime, and it allows to reproduce some well-known pN results.

Note that the Schwarzschild radial coordinate $r$ is an area coordinate; spheres with a radius $r=r_0$ have a surface area $4\pi r_0^2$, as can be read off from the metric \eqref{Eq_Schwarzschild_metric_SI}. The difference ${\Delta r := r_2 - r_1}$ is not the proper spatial distance between two events on the radial line on a ${t=\text{const.}}$ hypersurface.


\subsection{\label{SubSec_Geometry:pN} Post-Newtonian approximation}

Whenever the gravitational field, inside and in the neighborhood of a central object, is weak and all velocities are small compared to the speed of light, the pN framework is applicable. It is a method to solve Einstein's field equation to a given order of accuracy. For an overview of the pN framework, we recommend the books \citep{Kopeikin:Book:2011, Poisson:2014, Soffel:1989}, and the references therein. Modern conventions of the International Astronomical Union (IAU) and the International Earth Rotation and Reference Systems Service (IERS) use a first-order pN spacetime, see \citep{Soffel:2003} and \citep{IERS:2010}. 

For the first-order stationary pN approximation of a general relativistic spacetime outside the Earth, we have to use the metric 
\begin{subequations}
\label{Eq_pN_metric}
\begin{align}
	g_{00} &= - \left( 1 - \dfrac{2 U}{c^2} + \dfrac{2 U^2}{c^4} \right) + \mathcal{O}(c^6) \, , \\
	g_{0i} &= - \, \dfrac{4 U^i}{c^3} + \mathcal{O}(c^5) \, , \\
	g_{ij} &= \delta_{ij} \left( 1 + \dfrac{2U}{c^2} \right) + \mathcal{O}(c^4) \, ,
\end{align}
\end{subequations}
where the potentials $U,U^i$ satisfy the equations
\begin{subequations}
\begin{align}
	\label{Eq_pN_potentials}
	\Delta U(\mathbf{X}) &= - 4\pi G \rho(\mathbf{X}) \, ,\\
	\Delta U^i(\mathbf{X}) &= -4\pi G \rho^i(\mathbf{X}) = -4\pi G \rho(\mathbf{X}) \, v^i(\mathbf{X}) \, ,
\end{align}
\end{subequations}
and $\Delta$ is the usual Laplace operator. The energy (mass) density $\rho$ and the energy density flux $\rho^i$ are related to the energy-momentum tensor of the Earth by $\rho = (T^{00} + T^{ii})/c^2$ and $\rho^i = T^{0i}/c$, evaluated in the Geocentric Celestial Reference System (GCRS) with Cartesian coordinates $(T,X,Y,Z)$, and $v^i$ is the gravitating matter's velocity. For the scalar and vector potentials, one obtains \citep{Chandrasekhar:1965}
\begin{subequations}
\begin{align}
	\label{Eq_pN_potentials2}
	U(X) &= G \int d^3 X' \, \dfrac{\rho(\mathbf{X}')}{|\mathbf{X}-\mathbf{X}'|} \, , \\
	U^i(X) &= G \int d^3 X' \, \dfrac{\rho(\mathbf{X}') \, v^i(\mathbf{X}')}{|\mathbf{X}-\mathbf{X}'|} \, .
\end{align}
\end{subequations}
To construct the pN approximation of the Schwarzschild spacetime, we have to use the Newtonian gravitational potential of a spherically symmetric mass distribution \footnote{Here, we adopt the positive sign convention, which is commonly used in geodesy. Note that in physics the attractive gravitational potential is usually taken to be negative.},
\begin{align}
	\label{Eq_pN_potential_sph}
	U = GM/R \, ,
\end{align}
where $R=\sqrt{X^2+Y^2+Z^2}$ is the distance to the center of mass in the GCRS. The vector potential $U^i$ vanishes identically because there are no mass currents present. Hence, the pN metric \eqref{Eq_pN_metric} for the first-order approximation of the Schwarzschild spacetime becomes
\begin{multline}
	\label{Eq_pN_metric_Schwarzschild}
	g = -\left( 1 - \dfrac{2GM}{c^2r} + \dfrac{2G^2M^2}{c^4 r^2} \right) c^2 \mathrm{d}T^2 \\
	+ \left( 1 + \dfrac{2GM}{c^2r} \right) (\mathrm{d}X^2 + \mathrm{d}Y^2 + \mathrm{d}Z^2) \, .
\end{multline}
Now, we introduce spherical coordinates $(R,\Theta,\Phi)$, by the usual relations to $(X,Y,Z)$, and rewrite the metric \eqref{Eq_pN_metric_Schwarzschild} in the new coordinates to obtain
\begin{multline}
	\label{Eq_pN_metric_Schwarzschild_spherical}
	g = -\left( 1 - \dfrac{2m}{R} + \dfrac{2m^2}{R^2} \right) c^2 \mathrm{d}T^2 \\
	+ \left( 1 + \dfrac{2m}{R} \right) (\mathrm{d}R^2 + R^2 \mathrm{d}\Theta^2 + R^2 \sin^2\Theta \mathrm{d}\Phi^2) \, ,
\end{multline}
where we use the relation between $M$ and $m$.


\subsection{\label{SubSec_Geometry:Coords} Radial coordinates}

For the pN approximation of the Schwarzschild metric, see Eq.\ \eqref{Eq_pN_metric_Schwarzschild_spherical}, an isotropic radial coordinate $R$ is used. Hence, the spatial part of the metric is conformally flat, i.e.\ it appears to be the Minkowski line element modified by a coordinate dependent factor. To compare the pN metric to the Schwarzschild metric \eqref{Eq_Schwarzschild_metric_SI}, we have to either transform the metric \eqref{Eq_pN_metric_Schwarzschild_spherical} to area coordinates, or to transform the metric \eqref{Eq_Schwarzschild_metric_SI} to isotropic coordinates. To do the latter, we must have
\begin{multline}
	\left( 1- \dfrac{2m}{r} \right)^{-1} \mathrm{d}r^2 + r^2 \mathrm{d}\vartheta^2 + r^2 \sin^2 \vartheta \, \mathrm{d}\varphi^2 \\
	= f(\lambda) \left( \mathrm{d}\lambda^2 + \lambda^2 \mathrm{d}\vartheta^2 + \lambda^2\sin^2 \vartheta \, \mathrm{d}\varphi^2 \right) \, ,
\end{multline}
where $\lambda(r)$ is the new isotropic Schwarzschild radial coordinate. This yields a differential equation for $\mathrm{d}r/\mathrm{d}\lambda$,
\begin{align}
	\left( 1- \dfrac{2m}{r} \right)^{-1/2} \dfrac{\mathrm{d}r}{r} = \dfrac{\mathrm{d}\lambda}{\lambda} \, ,
\end{align}
which is solved by
\begin{align}
	r = \lambda \left( 1 + \dfrac{m}{2\lambda} \right)^2 \, .
\end{align}
The Schwarzschild metric in isotropic coordinates finally reads
\begin{multline}
	\label{Eq_Schwarzschild_metric_iso}
	g = - \left( \dfrac{1-m/(2\lambda)}{1+m/(2\lambda)} \right)^2 c^2 \mathrm{d}t^2 + \left( 1 + \dfrac{m}{2\lambda} \right)^4 \left( \mathrm{d}\lambda^2 \right. \\
		\left. + \lambda^2 \mathrm{d}\vartheta^2 + \lambda^2 \sin^2 \vartheta \mathrm{d}\varphi^2 \right) \, .
\end{multline}
Here, $\epsilon := m/\lambda$ is a small quantity, and for the region outside the Earth's surface it is less than $10^{-9}$.
The pN metric can be obtained now by expanding the spatial part in Eq.\ \eqref{Eq_Schwarzschild_metric_iso} to first-order and $g_{00}$ to second-order in $\epsilon$. In this way, we indeed recover the pN metric \eqref{Eq_pN_metric_Schwarzschild_spherical} after the identification $\lambda \hat{=} R$ at the given level of accuracy.


\section{\label{Sec_EOM} Equations of motion}

In GR, massive test bodies move on timelike geodesics. These are solutions of the geodesic equation
\begin{align}
	\label{Eq_EOM_geodesic}
	\ddot{x}^\mu + \Gamma^\mu{}_{\nu\sigma} \, \dot{x}^\nu \dot{x}^\sigma = 0 \, .
\end{align}
The worldline of the object is described by $x^\mu(\tau)$, and the overdot denotes derivatives w.r.t.\ the proper time $\tau$. The proper time is defined by the normalization of the four-velocity $u = \dot{x}$ according to
\begin{align}
	\label{Eq_propertime_def}
  	g(u,u) = g_{\mu\nu} u^\mu u^\nu = -c^2 \, .
\end{align}
The Christoffel symbols $\Gamma^\mu{}_{\nu\sigma}$ can be calculated from the metric by
\begin{align}
	\Gamma^\mu{}_{\nu\sigma} = \dfrac{1}{2} g^{\mu\lambda} \left( \partial_\nu g_{\sigma\lambda} + \partial_\sigma g_{\lambda\nu} - \partial_\lambda g_{\nu\sigma} \right) \, .
\end{align}
The EOM can also be derived using, e.g., the Lagrange or the Hamilton formalism, in which the Lagrangian for the motion of point-like test bodies is 
\begin{align}
	\label{Eq_general_Lagrangian_def}
	2 \mathcal{L} := g_{\mu\nu} \dot{x}^\mu \dot{x}^\nu \, .
\end{align}

We now introduce a general framework to derive the EOM that covers all cases which we are going to discuss in the following. The EOM for the Schwarzschild spacetime and its pN approximation are then contained as special examples. Therefore, we introduce the general metric
\begin{align}
	\label{Eq_EOM_metric}
	g = g_{tt}(\xi)\, \mathrm{d}t^2 + g_{11}(\xi)\, \mathrm{d}r^2 + g_{22}(\xi)\, \mathrm{d}\vartheta^2 + g_{33}(\xi,\vartheta)\, \mathrm{d}\varphi^2 \, ,
\end{align}
where we use $\xi$ as a radial coordinate, and angles $\vartheta, \varphi \, \in \, S_2$. The results derived in the following are valid for the Schwarzschild spacetime, in this case $\xi$ is either identified with the area coordinate $r$, or with the isotropic coordinate $\lambda$. For the first-order pN approximation of the Schwarzschild spacetime, $\xi$ is identified with the geocentric radial coordinate $R$. The metric functions must be chosen accordingly for either case. Note that, according to our sign convention, $g_{00}<0$, and all $g_{ij} > 0$. The Lagrangian for the motion in the spacetime \eqref{Eq_EOM_metric} is
\begin{align}
	\label{Eq_EOM_Lagrangian}
  	2 \mathcal{L} &= g_{tt}(\xi)\, \dot{t}^2 + g_{11}(\xi)\, \dot{r}^2 + g_{22}(\xi)\, \dot{\vartheta}^2 + g_{33}(\xi,\vartheta)\, \dot{\varphi}^2  \, .
\end{align}

The symmetry of the situation at hand allows to restrict the motion to the equatorial plane and all quantities in the following are assumed to be evaluated at $\vartheta = \pi/2$. 
The induced 3-dimensional metric in the equatorial plane is denoted as
\begin{align}
	\label{Eq_EOM_metric_equatorial}
	g^{(3)} = g_{tt}(\xi)\, \mathrm{d}t^2 + g_{11}(\xi)\, \mathrm{d}\xi^2 + g_{33}(\xi)\, \mathrm{d}\varphi^2 \, .
\end{align}

Since $\partial_t$ and $\partial_\varphi$ are Killing vector fields of the spacetime, there are two constants of motion, which are related to the energy $E$ and angular momentum $L$:
\begin{subequations}
\label{Eq_EOM_COM}
	\begin{align}
		E &:= - g_{tt}(\xi)\, \dot{t} \, ,\\
		L &:= g_{33}(\xi)\, \dot{\varphi} \, .
	\end{align}
\end{subequations}
Using the canonical conjugated momenta $p_\mu := \partial \mathcal{L} / \partial \dot{x}^\mu$ and the inverse metric components $g^{\mu\nu}$, we can also construct the Hamiltonian
\begin{align}
	\label{Eq_general_Hamiltonian_def}
	2\mathcal{H} = g^{\mu\nu} p_\mu p_\nu \, .
\end{align}
For the metric \eqref{Eq_EOM_metric_equatorial}, we obtain $(p_\vartheta = 0)$
\label{Eq_general_momenta}
\begin{align}
	\big( p_t, p_\varphi, p_\xi \big) = \big(-E, L, g_{11}(\xi)\, \dot{\xi} \big) \, .
\end{align}
Hence, two of the momenta are fixed by constants of motion, and the Hamiltonian (in the equatorial plane) becomes
\begin{align}
	\label{Eq_general_Hamiltonian}
	2 \mathcal{H} &= g^{tt}(\xi)\, p_t^2 + g^{11}(\xi)\, p_\xi^2 + g^{33}(\xi)\, p_\varphi^2 \notag \\
	&= g^{tt}(\xi)\, E^2 + g^{11}(\xi)\, p_\xi^2 + g^{33}(\xi)\, L^2 \, .
\end{align}


\subsection{\label{SubSec_EOM:properTime} Proper time parametrization}

The EOM, parameterized by the proper time of the respective test body, is given by, e.g., the Euler-Lagrange equations for the Lagrangian \eqref{Eq_EOM_Lagrangian}. For the azimuthal motion we obtain, see Eq.\ \eqref{Eq_EOM_COM},
\begin{align}
	\label{Eq_general_diffphi_pt}
	\dot{\varphi} = \dfrac{\mathrm{d}\varphi}{\mathrm{d}\tau} = \dfrac{L}{g_{33}(\xi)} \, .
\end{align}
The normalization of the four-velocity according to Eq.\ \eqref{Eq_propertime_def} gives a first-order differential equation for the radial motion, that is
\begin{align}
	\label{Eq_general_diffr_pt}
	\dot{\xi}^2 = -\dfrac{1}{g_{11}(\xi)} \left( c^2 + \dfrac{E^2}{g_{tt}(\xi)} + \dfrac{L^2}{g_{33}(\xi)} \right) \, ,
\end{align}
where we inserted the constants of motion from Eq.\ \eqref{Eq_EOM_COM}. Thereupon, we can construct a second-order differential equation by taking one more derivative with respect to proper time $\tau$,
\begin{align}
	\label{Eq_general_diff2r_pt}
	\ddot{\xi} = -\dfrac{1}{2} \diff{}{\xi} \left[ \dfrac{1}{g_{11}(\xi)} \left( c^2 + \dfrac{E^2}{g_{tt}(\xi)} + \dfrac{L^2}{g_{33}(\xi)} \right) \right] \, .
\end{align}
The EOM \eqref{Eq_general_diffphi_pt} and \eqref{Eq_general_diffr_pt} or \eqref{Eq_general_diff2r_pt} can now be solved using different methods. The initial conditions $(\xi_0, \varphi_0)$ and $(\dot{\xi}_0, \dot{\varphi}_0)$ at some $\tau = \tau_0$ must be specified. They are related to the constants of motion, as we will show in the next sections. Note that the advantage of Eq.\ \eqref{Eq_general_diff2r_pt} over the first-order equation \eqref{Eq_general_diffr_pt} is that it automatically takes care of all turning points along the orbit, which are, e.g., at the perigee and apogee of any bound orbit. Circular orbits can be found by equating Eqs.\ \eqref{Eq_general_diffr_pt} and \eqref{Eq_general_diff2r_pt} to zero at the same time.
 

\subsection{\label{SubSec_EOM:coordTime} Coordinate time parametrization}

To compare the solutions of the geodesic equation in the Schwarzschild spacetime to its pN approximation (and to Newtonian Kepler orbits) later on, we need an EOM that is parametrized by the coordinate time $t$. We can reparametrize all orbits using $\mathrm{d}\xi/\mathrm{d}t = \dot{\xi}/\dot{t}$ and $\mathrm{d}\varphi/\mathrm{d}t = \dot{\varphi}/\dot{t}$. We define derivatives w.r.t.\ the coordinate time $t$ by the symbol $\overset{\circ}{x} := {\mathrm{d}x/\mathrm{d}t}$. Thereupon, we obtain
\begin{subequations}
\label{Eq_general_diff_ct}
\begin{align}
	\overset{\circ}{\varphi} &:= \dfrac{\mathrm{d}\varphi}{\mathrm{d}t} = -\dfrac{L}{E} \dfrac{g_{tt}(\xi)}{g_{33}(\xi)} \, , \\
	\overset{\circ}{\xi}{}^2 &:= \left( \dfrac{\mathrm{d}\xi}{\mathrm{d}t} \right)^2 = - \dfrac{g_{tt}(\xi)}{g_{11}(\xi)} \left( \dfrac{c^2 g_{tt}(\xi)}{E^2} + \dfrac{L^2}{E^2} \dfrac{g_{tt}(\xi)}{g_{33}(\xi)} + 1 \right) \, .
\end{align}
\end{subequations}
Again, taking one more derivative with respect to coordinate time $t$, we get the second-order differential equation
\begin{align}
\label{Eq_general_diff_ct2}
	\overset{\circ \circ }{\xi} = -\dfrac{1}{2} \diff{}{\xi} \left[ \dfrac{g_{tt}(\xi)}{g_{11}(\xi)} \left( \dfrac{c^2 g_{tt}(\xi)}{E^2} + \dfrac{L^2}{E^2} \dfrac{g_{tt}(\xi)}{g_{33}(\xi)} + 1 \right) \right] \, .
\end{align}


\subsection{\label{SubSec_EOM:IC} Initial conditions}

We can relate the constants of motion $L$ and $E$ to orbital parameters. For the perigee $\xi_p$ and apogee $\xi_a$ of a bound orbit we introduce the well known formulae
\begin{align}
	\label{Eq_general_orbitalParam}
	\xi_p = (1-e) a \, , \quad \xi_a = (1+e) a \, ,
\end{align}
by which we define a semi-major axis $a$ and an eccentricity $e$ also in the relativistic setting. These radii mark the turning points of a bound orbit, $\dot{\xi} |_{\xi_p} = 0 = \dot{\xi} |_{\xi_a}$. Therefore, we obtain
\begin{subequations}
\begin{align}
	L^2 &= c^2 \dfrac{g_{tt}(\xi_a) - g_{tt}(\xi_p)}{ \dfrac{g_{tt}(\xi_p)}{g_{33}(\xi_p)}  - \dfrac{g_{tt}(\xi_a)}{g_{33}(\xi_a)} } \, , \\
	E^2 &= - g_{tt}(\xi_p) \left( c^2 + \dfrac{L^2}{g_{33}(\xi_p)} \right) \, .
\end{align}
\end{subequations}
Given an initial position $(\xi_0, \, \varphi_0)$, we can also derive the initial velocities w.r.t.\ proper time
\begin{subequations}
\begin{align}
	\dot{\xi}_0 &= \sqrt{- \dfrac{1}{g_{11}(\xi_0)} \left( c^2 + \dfrac{E^2}{g_{tt}(\xi_0)} + \dfrac{L^2}{g_{33}(\xi_0)} \right)} \, , \\
	\dot{\varphi}_0 &= \dfrac{L}{g_{33}(\xi_0)} \, ,
\end{align}
\end{subequations}
and w.r.t.\ coordinate time we get
\begin{subequations}
\begin{align}
	\overset{\circ}{\xi}_0 &= \sqrt{- \dfrac{g_{tt}(\xi_0)}{g_{11}(\xi_0)} \left( \dfrac{c^2g_{tt}(\xi_0)}{E^2} + \dfrac{L^2}{E^2} \dfrac{g_{11}(\xi_0)}{g_{33}(\xi_0)} + 1 \right)} \, , \\
	\overset{\circ}{\varphi}_0 &= -\dfrac{L}{E} \dfrac{g_{tt}(\xi_0)}{g_{33}(\xi_0)} \, .
\end{align}
\end{subequations}
Note that $g_{tt}$ is negative such that there is no problem with the square root.


\subsection{\label{SubSec_EOM:Schwarzschild} Orbits in the Schwarzschild spacetime}

In the following, we apply the general results of the previous section to the Schwarzschild spacetime. Since the relativistic EOM is usually considered in area coordinates and a proper time parametrization, we do only explicitly give this result. However, the other cases can be easily deduced from Eqs.\ \eqref{Eq_general_diff_ct} and \eqref{Eq_general_diff_ct2}.

Written in the form of Eq.\ \eqref{Eq_EOM_metric_equatorial}, where we identify $\xi = r$, the induced Schwarzschild metric in the equatorial plane reads
\begin{subequations}
\begin{align}
	g_{00} &= - A(r)\,c^2 \, , \\
	g_{11} &= A(r)^{-1} \, , \\
	g_{33} &= r^2 \, .
\end{align}
\end{subequations}
The Lagrangian for a particle moving in the Schwarzschild spacetime now becomes
\begin{align}
	\label{Eq_Schwarzschild_Lagrangian}
  	2 \mathcal{L} = - A(r)\, c^2 \dot{t}^2 + A(r)^{-1}\, \dot{r}^2 + r^2 \dot{\varphi}^2 \, ,
\end{align}
and the two constants of motion, related to energy $E$ and angular momentum $L$, are given by
\begin{subequations}
\begin{align}
	\label{Eq_Schwarzschild_com}
  	E &= c^2 A(r)\, \dot{t} \, , \\
  	L &= r^2 \dot{\varphi} \, .
\end{align}
\end{subequations}
The canonical conjugated momenta become
	\label{Eq_Schwarzschild_momenta}
\begin{align}
  	\big(p_t, p_\varphi, p_r \big) = \big(-E,L,A(r)^{-1} \dot{r} \big) \, , 
\end{align}
and the Hamiltonian reads
\begin{align}
	\label{Eq_Schwarzschild_Hamiltonian}
  	2\mathcal{H} = -\dfrac{p_t^2}{A(r)~c^2} + A(r) \, p_r^2 + \dfrac{p_\varphi^2}{r^2} \, .
\end{align}
The motion of the test body is described by
\begin{subequations}
\label{Eq_Schwarzschild_EOM}
\begin{align}
	\dot{\varphi} &= \dfrac{L}{r^2} \quad \Rightarrow \quad \ddot{\varphi} = -\dfrac{2\dot{r}\dot{\varphi}}{r} \\
	\dot{r}^2 &= - A(r)^{-1} \left( c^2 - \dfrac{E^2}{c^2 A(r)} + \dfrac{L^2}{r^2} \right) \\
	\ddot{r} &= -\dfrac{1}{2} \diff{}{r} \left[ A(r)^{-1} \left( c^2 - \dfrac{E^2}{c^2 A(r)} + \dfrac{L^2}{r^2} \right) \right] \, .
\end{align}
\end{subequations}

The constants of motion are related to the initial conditions by
\begin{subequations}
	\label{Eq_Schwarzschild_initcond}
	\begin{align}
	\dot{\varphi}_0 &= \dfrac{L}{r_0^2} \, , \\
  	\dot{r}_0 &= \sqrt{- A(r_0) \left( c^2 - \dfrac{E^2}{c^2 A(r_0)} + \dfrac{L^2}{r_0^2} \right)} \, ,
	\end{align}
\end{subequations}
and the relations to the orbital elements $(r_p, r_a)$ are
\begin{subequations}
\label{Eq_Schwarzschild_orbitalparam}
\begin{align}
	\dfrac{L^2}{c^2} &= \dfrac{ A(r_p) - A(r_a) }{ \frac{A(r_a)}{r_a^2} - \frac{A(r_p)}{r_p^2} } \, , \\
	\dfrac{E^2}{c^2} &= \left( \dfrac{ L^2 }{ r_a^2 } + c^2 \right) A(r_a) \, .
\end{align}
\end{subequations}


\subsection{\label{SubSec_EOM:pN} Orbits in the post-Newtonian approximation}

In this section, we apply the general results to the first-order pN approximation of the Schwarzschild spacetime. In the pN framework, orbits are usually parameterized by the coordinate time, and the isotropic radial coordinate is used. Doing so, the pN EOM appears to look like a relativistically modified Newtonian EOM.

In the form \eqref{Eq_EOM_metric_equatorial}, where we identify $\xi$ with $R$, the induced pN metric in the equatorial plane is given by
\begin{subequations}
\begin{align}
	g_{00} &= -c^2 \left( 1-\dfrac{2m}{R}+\dfrac{2m^2}{R^2} \right) \, , \\
	g_{11} &= \left( 1+\dfrac{2m}{R} \right) \, , \\
	g_{33} &= r^2 \left( 1+\dfrac{2m}{R} \right) \, .
\end{align}
\end{subequations}
We use the small parameter $\epsilon = m/R$ for order counting. Being consistent to $\mathcal{O}(\epsilon)$, the EOM can be written as the Keplerian orbital equation with pN correction terms \footnote{The first-order pN approximation is usually defined by order counting in the EOM. Terms of order $\mathcal{O}(\epsilon^n)$ are proportional to $c^{-2n}$ and said to be at the $n$-th pN order.}
\begin{subequations}
\label{Eq_pN_EOM}
\begin{align}
	\overset{\circ \circ}{R} &= -\dfrac{GM}{R^2} + R \overset{\circ}{\varphi}{}^2 + \dfrac{1}{c^2} \dfrac{GM}{R^3} \big( 4GM + 3R \overset{\circ}{R}{}^2 - R^3 \overset{\circ}{\varphi}{}^2 \big) \\
	\overset{\circ \circ}{\Phi} &= -2 \dfrac{\overset{\circ}{R}\overset{\circ}{\Phi}}{R} \left( 1-\dfrac{1}{c^2}\dfrac{2GM}{R} \right) \, .
\end{align}
\end{subequations}
The well-known Keplerian EOM is recovered in the limit $\epsilon \to 0$, i.e. $c \to \infty$. Introducing the Cartesian position vector $\mathbf{x} = (X,Y,Z)$ in the GCRS, the two equations can be combined to
\begin{multline}
\label{Eq_pN_EOM_Cartesian}
	\overset{\circ \circ}{\mathbf{x}} = -\dfrac{GM}{R^3}\mathbf{x} \\ 
	+ \dfrac{1}{c^2} \dfrac{GM}{R^3} \left[\left(\frac{4GM}{R}-\overset{\circ}{\mathbf{x}} \cdot \overset{\circ}{\mathbf{x}}\right)\mathbf{x} + 4\left(\mathbf{x} \cdot \overset{\circ}{\mathbf{x}}\right) \overset{\circ}{\mathbf{x}}\right] \, ,
\end{multline}
where $R = ||\mathbf{x}||_2$. This equation is a special case of
\begin{multline}
\label{Eq_pN_EOM_Cartesian_general}
	\overset{\circ \circ}{\mathbf{x}} = \nabla U + c^{-2} \big( -4 U \nabla U -4\big(\nabla U \cdot \overset{\circ}{\mathbf{x}}\big) \overset{\circ}{\mathbf{x}} + \big(\overset{\circ}{\mathbf{x}} \cdot \overset{\circ}{\mathbf{x}} \big) \nabla U  \\
	-4 \overset{\circ}{\mathbf{x}} \times (\nabla \times \mathbf{U}) \big) \, ,
\end{multline}
which is valid for a general potential $U$ and also includes gravitomagnetic (Lense-Thirring) effects caused by the vector potential $\mathbf{U}$, see, e.g., \citep{Soffel:1989, Koop:1993} and references therein.

\section{\label{Sec_SolMethods} Test orbits and solution methods}

In the following, we consider different methods to solve the EOM in the Schwarzschild spacetime and its first-order pN approximation. We use a set of satellite test orbits with different eccentricities and altitudes to test the solution methods and to quantify their respective accuracy by comparison to the analytical reference solution.

 
\subsection{\label{SubSec_SolMethods:testOrbits} Set of test orbits}

To compare the different solution methods for the EOM, we consider the orbits shown in Tab.\ \ref{Tab_orbits}. All orbits are assumed to lie in the equatorial plane, $\vartheta = \pi/2$, such that the orbit modeling reduces to a 2-dimensional problem. All values for orbital elements such as the semi-major axis or the perigee are taken to be defined using the Schwarzschild area coordinate $r$.
\begin{table}[]
\centering
	\begin{tabular}{c|c c c}	
	\hline \hline
 	orbit & shape, type & eccentricity & semi-major axis [m]\\
  	\hline 
  	$\#1$ & c, MEO & 0 & $2.79776 \cdot 10^7$ \\
  	$\#2$ & e, MEO & 0.162 & $2.79776 \cdot 10^7$ \\
  	$\#3$ & e, MEO & 0.300 & $2.79776 \cdot 10^7$ \\
  	$\#4$ & e, MEO & 0.450 & $2.79776 \cdot 10^7$ \\
  	$\#5$ & e, MEO & 0.600 & $2.79776 \cdot 10^7$ \\
  	$\#6$ & e, MEO & 0.750 & $2.79776 \cdot 10^7$ \\
  	$\#7$ & e, LEO & 0.2 & $8.5\, \cdot 10^6$ \\
  	$\#8$ & e, LEO & 0.001 & $6.8\, \cdot 10^6$ \\
  	\hline \hline
	\end{tabular}
	\caption{\label{Tab_orbits} Orbital parameters for the eight different test orbits. These orbits are used to compare the different solution methods for the EOM. The shape of the orbit is either circular (c) or elliptical (e), and we use Low Earth Orbits (LEO) as well as Medium Earth Orbits (MEO). The orbital parameters (perigee, apogee) are defined in Eq.\ \eqref{Eq_Schwarzschild_orbitalparam}, and the semi-major axis as well as the eccentricity are defined in Schwarzschild area coordinates, see Eq.\ \eqref{Eq_general_orbitalParam} with $\xi \equiv r $.}
\end{table}

Without loss of generality, we can assume that all considered orbits start at their respective perigee, such that $r_0 = r_p$. Hence, initially $\dot{r}_0 = 0$ for all cases. Furthermore, we assume initially $\phi_0 = 0$ to fix the argument of the perigee. 

Orbit $\#1$ is circular with a radius {$r_0 = 27977600\,$m}, which is the altitude of the Galileo satellites. Orbit $\#2$ is characterized by a small eccentricity and corresponds roughly to the orbit of one of the Galileo satellites 5 and 6, which were not successfully launched into a circular orbit. The remaining orbits $\#3$ -- $\#6$ have larger eccentricities but we keep the same semi-major axis. Orbit $\#7$ is a low Earth orbit (LEO) with a moderate eccentricity, whereas orbit $\#8$ is an almost circular LEO at very low altitude. Choosing these orbits, we aim to cover a broad range of possible scenarios to test the different solution methods.

For the equatorial radius of the Earth \footnote{Since we consider a spherically symmetric gravitational field, the Earth is modeled as a sphere.}, we use $r_\oplus = 6378137 \,$m. From the eccentricity and semi-major axis, see Table \ref{Tab_orbits}, the initial angular velocity and the constants of motion can be calculated using Eqs.\ \eqref{Eq_Schwarzschild_initcond} and \eqref{Eq_Schwarzschild_orbitalparam}. These initial conditions are kept the same for all solution methods that we describe in the following.

Note that we use the terms semi-major axis and eccentricity here as defined by the area coordinate $r$ in the Schwarzschild spacetime, i.e. as defined by Eq.\ \eqref{Eq_Schwarzschild_orbitalparam} and Eq.\ \eqref{Eq_general_orbitalParam}, where $\xi \equiv r $. Hence, they do not fully coincide with (post-)Keplerian orbital elements. To have the same initial conditions for the pN orbits, the area coordinate needs to be transformed to an isotropic radial coordinate, and the initial angular velocity needs to be given w.r.t.\ the coordinate time, see Sec.\ \ref{Sec_accuracypN}.


\subsection{\label{SubSec_SolMethods:methods} Solution methods}

\subsubsection{\label{SubSubSec_SolMethods:methods:analytical} Analytical solution}

The analytical solution of the EOM can be given in terms of elliptic functions. Using $\mathrm{d}r/\mathrm{d}\varphi = \dot{r}/\dot{\varphi}$ yields a differential equation for $r(\varphi)$ that can be solved in terms of the Weierstrass elliptic function $\wp$ \citep{Hagihara:1930}. The solution is given by
\begin{align}
	\label{Eq_WeierstrassSolution}
	r(\varphi) = \dfrac{m}{2 \wp (\varphi - \varphi_{\text{in}}) + 1/6} \, ,
\end{align}
where $\varphi_{\text{in}}$ is related to the initial conditions according to
\begin{align}
	\varphi_{\text{in}} = \varphi_0 + \int_{y_0}^{\infty} \dfrac{\mathrm{d}z}{\sqrt{4z^3-g_2z-g_3}} \, , \, y_0 = \dfrac{1}{2}\left(\dfrac{m}{r_0} - \dfrac{1}{6} \right) \, .
\end{align}
The Weierstrass invariants $g_2$ and $g_3$ are determined by the constants of motion as follows:
\begin{subequations}
	\begin{align}
		g_2 &= \dfrac{1}{12} - \dfrac{c^2 m^2}{L^2} \, , \\
		g_3 &= \dfrac{1}{216} - \dfrac{1}{12}\dfrac{c^2 m^2}{L^2} - \dfrac{1}{4}\dfrac{m^2}{L^2}\left(\dfrac{E^2}{c^2}-c^2\right) \, .
	\end{align}
\end{subequations}
For details on the analytic solution and possible applications, we refer the reader to the seminal paper by Hagihara \citep{Hagihara:1930} and the work in \citep{Hackmann:2008a}, \citep{Hackmann:2008b}, and \citep{Hackmann:2009}. 

The analytical solution serves as the reference to check the accuracy of all other solution methods. If a solution in terms of $r(\tau),~ \varphi(\tau)$ is obtained, Eq.\ \eqref{Eq_WeierstrassSolution} can be used to calculate the actual value of the radius $r$ for a given value of the azimuthal angle $\varphi$. Thereupon, we can calculate the deviation from the analytical solution.


\subsubsection{\label{SubSubSec_SolMethods:methods:numerical} Numerical solution}

For the numerical integration of the EOM, we solve Eqs.\ \eqref{Eq_Schwarzschild_EOM} with a working precision of 32 digits using a Runge-Kutta (RK) integrator and equidistant proper time values $\tau_i$. The same numerical grid is used for the semi-analytical Lie-series method in the following and is specified in the corresponding section. Here, we use the numerical RK integrator that is implemented in the Mathematica computer algebra system.


\subsubsection{\label{SubSubSec_SolMethods:methods:Lie} Lie series approach}

The semi-analytical Lie series approach is based on the Hamiltonian formulation of the EOM. The Lie-series formalism was applied to Newtonian orbital mechanics by Lelgemann \citep{Lelgemann:1983} as a tool to construct a fully analytical theory of motion based on canonical variables and Lie transformations, see also \citep{Deprit:1969,Hori:1973}. The approach was developed towards a second-order analytical orbital theory by Cui \citep{Cui:1997}. In \citep{Mai:2014}, it is shown how a Lie series approach can be used for semi-analytical numerical orbit integration, and how it can be improved by using parallel computing techniques for the series coefficient calculation. Since the Lie-series approach has proven to be very useful in Newtonian dynamics, we test the method also for the relativistic case. Here, we base our considerations on the results for the semi-analytical orbit integrator, where the time is the expansion variable. Therefore, we use the Hamiltonian \eqref{Eq_Schwarzschild_Hamiltonian} that generates the EOM. For the Lie series, we recursively define coefficients $f^{\mu}_{(k)}$ and $h_{\mu, (k)}$ by the Poisson brackets
\begin{subequations}
\begin{align}
	f^{\mu}_{(k+1)} &= \partial_\tau f^{\mu}_{(k)} + \{ f^{\mu}_{(k)}, \mathcal{H} \} \, , \\
	h_{\mu, (k+1)} &= \partial_\tau h_{\mu, (k)} + \{ h_{\mu, (k)}, \mathcal{H} \} \, ,
\end{align}
\end{subequations}
where $\mu$ is a spacetime index, labeling the coordinates and the components of the canonical momenta, and $k$ gives the order of the Lie-series approximation. For the multi-dimensional Poisson bracket on the phase space, see, e.g., \citep{Mai:2014}, we have
\begin{subequations}
\begin{align}
	\{ f^{\mu}_{(k)}, \mathcal{H} \} &=  \dfrac{\partial f^{\mu}_{(k)}}{\partial x^\nu} \dfrac{\partial \mathcal{H}}{\partial p_\nu} - \dfrac{\partial f^{\mu}_{(k)}}{\partial p_\nu} \dfrac{\partial \mathcal{H}}{\partial x^\nu} \, , \\ 
	\{ h_{\mu, (k)}, \mathcal{H} \} &= \dfrac{\partial h_{\mu,(k)}}{\partial x^\nu} \dfrac{\partial \mathcal{H}}{\partial p_\nu} - \dfrac{\partial h_{\mu,(k)}}{\partial p_\nu} \dfrac{\partial \mathcal{H}}{\partial x^\nu} \, .
\end{align}
\end{subequations}
Note that $f^{\mu}_{(k)}$ and $h_{\mu, (k)}$ are functions of the phase space coordinates. The initial conditions for the recursive definitions are given by
\begin{subequations}
\begin{align}
	f^{\mu}_{(0)} &= x^\mu \, , \\
	h_{\mu,(0)} &= p_\mu \, .
\end{align}
\end{subequations}
Hence, all the coefficients $f^{\mu}_{(k)}$ and $h_{\mu,(k)}$ can be calculated by nested Poisson brackets. Thereupon, the solution of the EOM is obtained by the Lie-series
\begin{subequations}
\label{Eq_LieSeries}
\begin{align}
	x^\mu(\tau_0+\Delta \tau) &= \sum_{k=0}^\infty \dfrac{(\Delta \tau)^k}{k!} f^{\mu}_{(k)} |_{\tau_0} \, , \\
	p_\mu(\tau_0+\Delta \tau) &= \sum_{k=0}^\infty \dfrac{(\Delta \tau)^k}{k!} h_{\mu,(k)} |_{\tau_0} \, .
\end{align}
\end{subequations}

Introducing a finite upper limit of summation $k_{\text{max}}$ gives an approximation of the full series. A step size $\Delta \tau$ needs to be chosen, and the summation is then performed numerically in an iterative way starting with the initial point in phase space and calculating the next point by evaluating the series up to $k_{\text{max}}$. The new point in phase space is then used to read off the initial conditions for the next step. Therefore, the entire approach is said to be semi-analytic. The advantage over the direct numerical integration of the EOM is given by the expression of all series coefficients as analytical functions of the phase space coordinates. Therefore, the method may prove useful in the spectral domain, where individual terms can be studied and their impact on the orbit, as well as relativistic effects, might be quantified in detail.


\subsubsection{\label{SubSubSec_SolMethods:methods:XHPS} The XHPS integrator}

The XHPS is designed to simulate multi-satellite missions incorporating gravity and also non-gravitational perturbations using non-relativistic Newtonian mechanics. Its modular design allows to easily set up simulations at all levels of detail. Different numeric integration schemes are implemented, using, e.g., Runge-Kutta and multistep integrators. Due to the C++ code basis, the GNU MPFR library \citep{MPFR:2007} with variable-precision data types can be utilized, so that results can in general reach every desired numerical accuracy.

The XHPS uses the International Celestial Reference System (ICRS) \citep{Arias:1995}, and the International Terrestrial Reference System (ITRS) \citep{Boucher:2000} as Cartesian coordinate systems derived from the BCRS and GCRS definitions. These systems are realized according to the latest IERS conventions \citep{IERS:2010}, and in these reference systems, relativistic effects can be considered as pN corrections in the EOM. The pN correction term for a spherically symmetric gravitational field, see Eqs.\ \eqref{Eq_pN_EOM_Cartesian}, in Cartesian GCRS coordinates is
\begin{align}
	\label{Eq_pN_cart}
	\mathbf{a}_{\text{pN}} =  \frac{GM}{c^2 R^3} \left[\left(\frac{4GM}{R}-\overset{\circ}{\mathbf{x}} \cdot \overset{\circ}{\mathbf{x}} \right)\mathbf{x} + 4\left(\mathbf{x} \cdot \overset{\circ}{\mathbf{x}}\right)\overset{\circ}{\mathbf{x}}\right] \, .
\end{align}
See \citep{IERS:2010} and references therein for an overview of the IERS conventions that are used for the XHPS.
Here, the vector $\mathbf{x} = (X,Y,Z)$ is the three component Cartesian position vector in the GCRS and $\overset{\circ}{\mathbf{x}}$ its coordinate time derivative. The acceleration $\mathbf{a}_{\text{pN}}$ can be added to all other gravitational (Earth, Sun and other bodies) and non-gravitational (environmental) accelerations in the EOM that are implemented in the XHPS.

For all orbit calculations, the data type was set to 64 digits, and the 8th order Dormand-Prince RK integrator was used with a constant step size, resulting in an absolute numerical precision of at least 25 digits, or about $10^{-17}\,$m. 

As mentioned already at the beginning of this section, the orbital parameters in Tab.\ \ref{Tab_orbits} are defined using the Schwarzschild area coordinate $r$ and have to be transformed to be used in the (post-)Newtonian framework. For the pN orbits, the XHPS needs as initial conditions the angular velocity w.r.t.\ the coordinate time, $\overset{\circ}{\Phi}$, and the initial radius $R_0 = R_p$ (perigee) in isotropic coordinates. These are related to $\dot{\varphi}_0$ and $r_0$ by
\begin{align}
		\overset{\circ}{\Phi}_0 = \dot{\varphi}_0 \, \dfrac{c^2 A(r_0)}{r_0^2} \, , \quad r_0 = R_0 \left( 1 + \dfrac{m}{2R_0} \right)^2 \, .
\end{align}


\section{\label{Sec_Comparison} Comparison of different methods}


\subsection{\label{SubSec_Comparison:numerical} Numerical integration}

\begin{figure}
	\centering
	\includegraphics[width=0.49\textwidth]{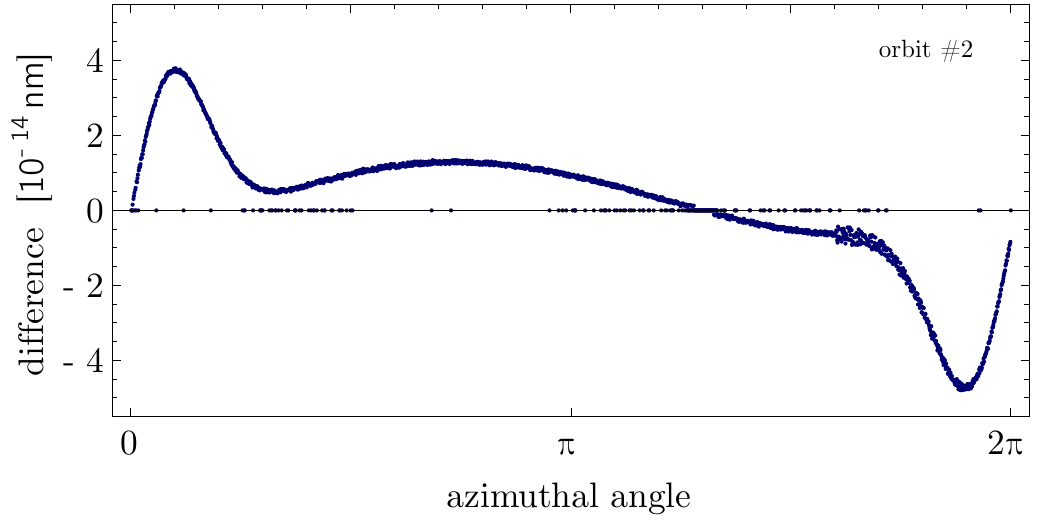}\\[5pt]
	\includegraphics[width=0.49\textwidth]{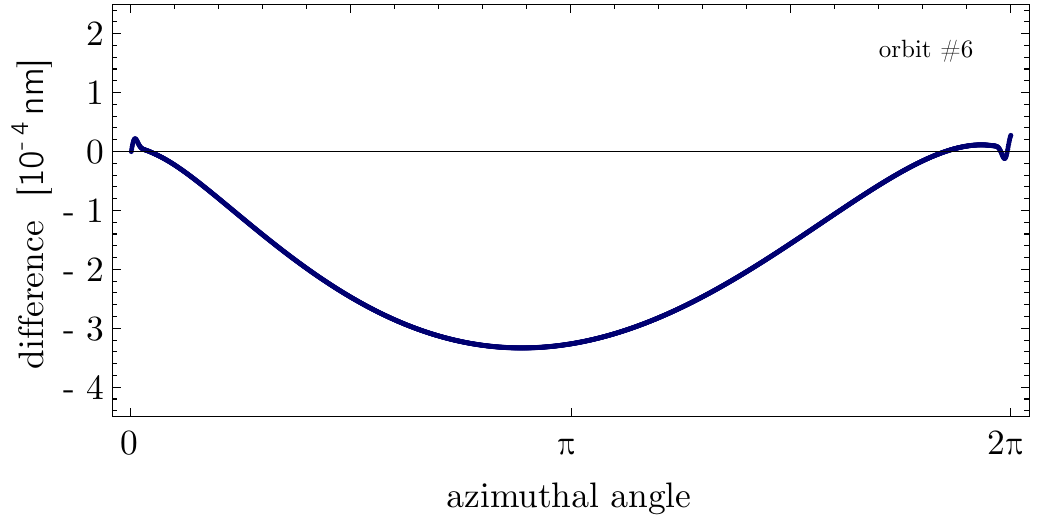}
	\caption{\label{Fig_num} The figure shows the accuracy of the numerical integration using a simple Runge-Kutta integrator, which is implemented in the computer algebra system {Mathematica}. We used a fixed grid of 2001 equidistant grid points and a working precision of 32 digits. We show the radial deviations from the analytical solution for orbit \#2 (top) and orbit \#6 (bottom). The values lying on the horizontal axis in the upper plot are significantly smaller than the scaling, i.e.\ orders of magnitude below $10^{-23}\,$m. Hence, for the given 32 digit precision, they are numerically effectively zero.}
\end{figure}

Figure \ref{Fig_num} shows the difference between the numerical orbit integration of Eq.\ \eqref{Eq_Schwarzschild_EOM} and the analytical solution for orbits $\#2$ and $\#6$, i.e.\ for the orbits with the smallest and largest eccentricity, respectively. With a working precision of 32 digits, the numerical orbits where obtained using a simple Runge-Kutta integrator on a fixed grid with 2001 equidistant grid points along the orbit. We used the computer algebra system Mathematica and the implemented functions for integrating second-order differential equations to obtain the solutions for $r(\tau)$ and $\varphi(\tau)$. For all considered cases, see Tab.\ \ref{Tab_orbits}, the accuracy of this numerical approach is in the sub-nanometer regime along one full orbital arc. 

The difference to the analytical solution is maximal for orbit $\#6$, which has the largest eccentricity of $\epsilon = 0.75$. But even in this case, the deviation is below $10^{-3}\,$nm. Hence, direct numerical integration of the EOM, with the given settings, yields an accuracy in the sub-nanometer regime after only a few seconds of computation time for one full orbit, and it is well suited to integrate the geodesic equation for high-precision results. Here, the purpose was to test and verify the applicability of the direct numerical integration method. Increasing the working precision and refining the numerical grid yields even more accurate orbits. However, to date the experimental capabilities do not allow for a more precise orbit validation, and the state of the art accuracy is achieved by laser ranging measurements in the (sub-)centimeter regime \citep{Sosnica:2018}.

\subsection{\label{SubSec_Comparison:Lie} Lie-series approach}

All test orbits are constructed using the semi-analytical Lie-series approach with a maximal order $k_{\text{max}} = 9$ and $k_{\text{max}} = 12$, respectively, which yield (sub-)nanometer accuracy for all cases.

The step size $\Delta \tau$ is chosen to be the same as for the direct numerical integration of the EOM. We have chosen an equidistant grid of 2001 points with a maximal integration time of $\tau_{\text{end}} = 4.67 \cdot 10^7\,$s for the orbits $\#1$ to $\#6$, while $\tau_{\text{end}}= 7.81 \cdot 10^3\,$s and $\tau_{\text{end}}= 5.57 \cdot 10^7\,$s for orbits $\#7$ and $\#8$, respectively. The time of integration is chosen such that in each case at least one full revolution is obtained.

To summarize the results for the semi-analytical Lie-series approach, we can say that for all considered test orbits the deviation from the analytical solution is in the (sub-)nanometer regime, with an upper bound for the Lie-series order at ${k_{\text{max}} = 9}$. For ${k_{\text{max}} = 12}$, this difference is at least three orders of magnitude smaller, i.e.\ the method becomes better with increasing order. Note, however, that there is no proof of convergence for the Lie-series \eqref{Eq_LieSeries}, see also Sec.\ VII in \citep{Tessmer:2013}.

\begin{figure}
	\centering
	\includegraphics[width=0.49\textwidth]{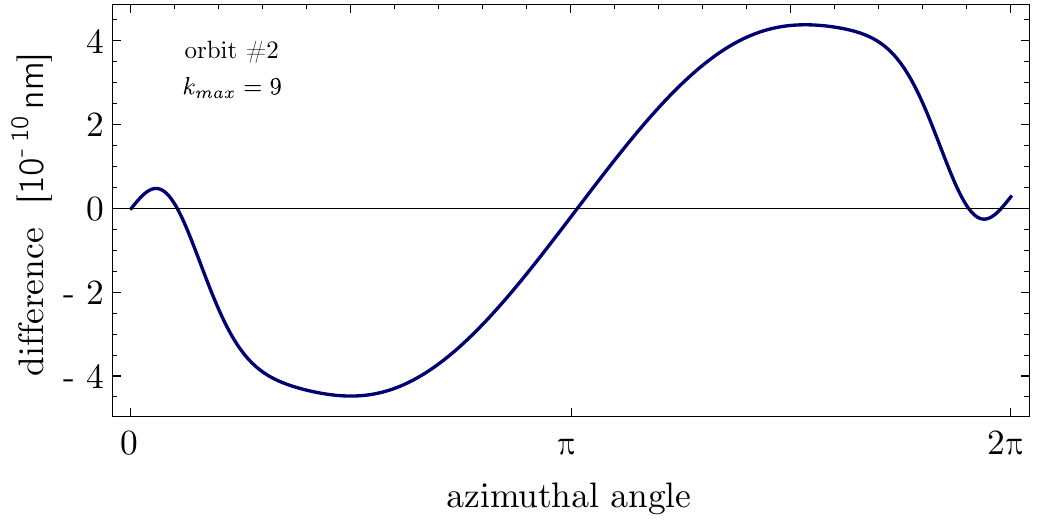}\\[2pt]
	\includegraphics[width=0.49\textwidth]{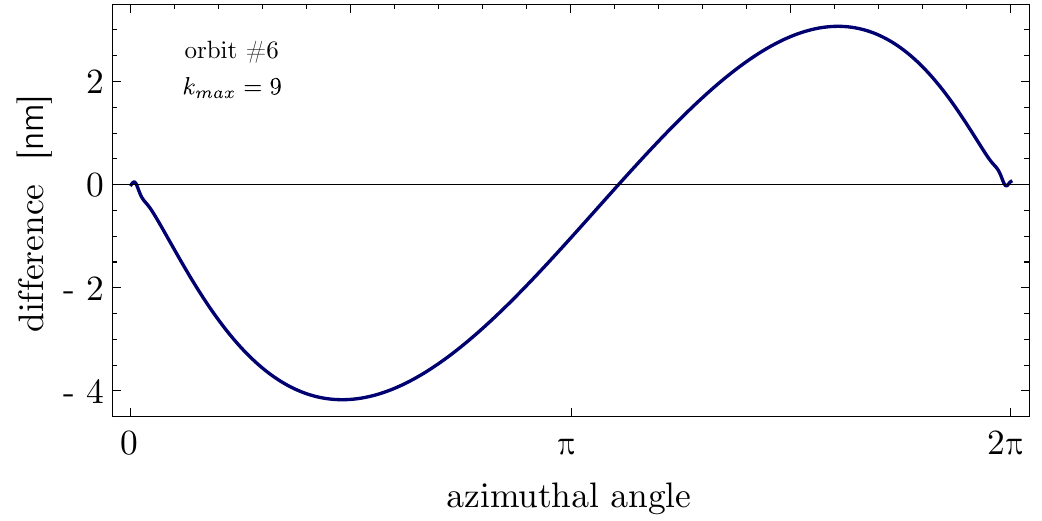}
	\caption{\label{Fig_lie} The figure shows the accuracy of the semi-analytical Lie-series approach for a maximal order $k_{\text{max}} = 9$. We show the radial deviations from the analytical solution for orbit \#2 (top) and orbit \#6 (bottom). We used a fixed grid of 2001 equidistant grid points and a working precision of 32 digits.}
\end{figure}

In Fig.\ \ref{Fig_lie}, we show the results for $k_{\text{max}} = 9$ and orbit $\#2$ as well as orbit $\#6$, to depict two exemplary results. The maximal deviation from the analytical solution is found for orbit $\#6$, and its magnitude is about $\pm 4\,$nm. For all other cases, the deviation is found to be orders of magnitude smaller. Hence, the semi-analytical Lie-series approach appears to work very well for solving the relativistic EOM using the Hamiltonian of the system and nested Poisson brackets. One drawback is the much larger computation time (minutes) compared to the direct numerical integration, which delivers results at the same level of accuracy. However, all Lie series coefficients are known analytically and their influence on the orbit and relativistic effects will be studied in a separate paper.

\section{\label{Sec_accuracypN} Accuracy of the post-Newtonian approximation}

To access the accuracy of the first-order pN approximation of the relativistic EOM, we have implemented the pN correction term ${ \mathbf{a}_{\text{pN}}}$, see Eq.\ \eqref{Eq_pN_cart}, into the XHPS. Since we analyze the spherically symmetric situation, we specify the Newtonian gravitational potential to be the pure monopole potential given by Eq.\ \eqref{Eq_pN_potential_sph}. In general, the XHPS can model detailed satellite properties and the coupling to various gravitational and environmental perturbations. Here, we simplify matters and test the accuracy of the first-order pN approximation w.r.t.\ the general relativistic solution. Therefore, the satellite is modeled as a point mass, which allows for neglecting effects due to a possible spinning or tumbling.

The reason to use the XHPS to investigate the pN orbits is twofold: this approach allows to i) test the accuracy of the first-order pN approximation by comparing the orbit to the solution of the geodesic equation in GR, and ii) to compare the acceleration caused by the pN correction terms to various disturbing forces, such as solar radiation pressure, at a later stage, see the next section.

\begin{figure}
	\centering
	\includegraphics[width=0.49\textwidth]{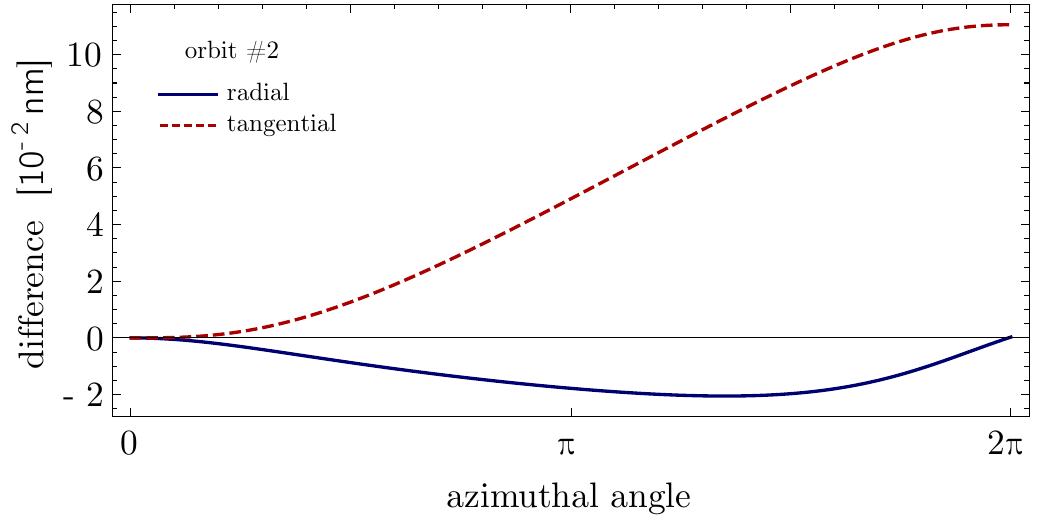}\\[2pt]
	\includegraphics[width=0.49\textwidth]{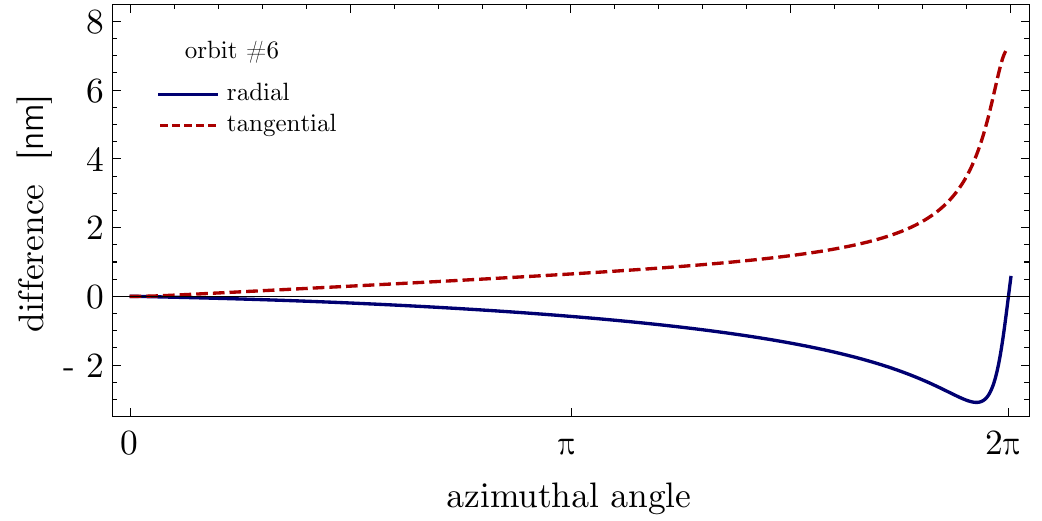}\\[2pt]
	\includegraphics[width=0.49\textwidth]{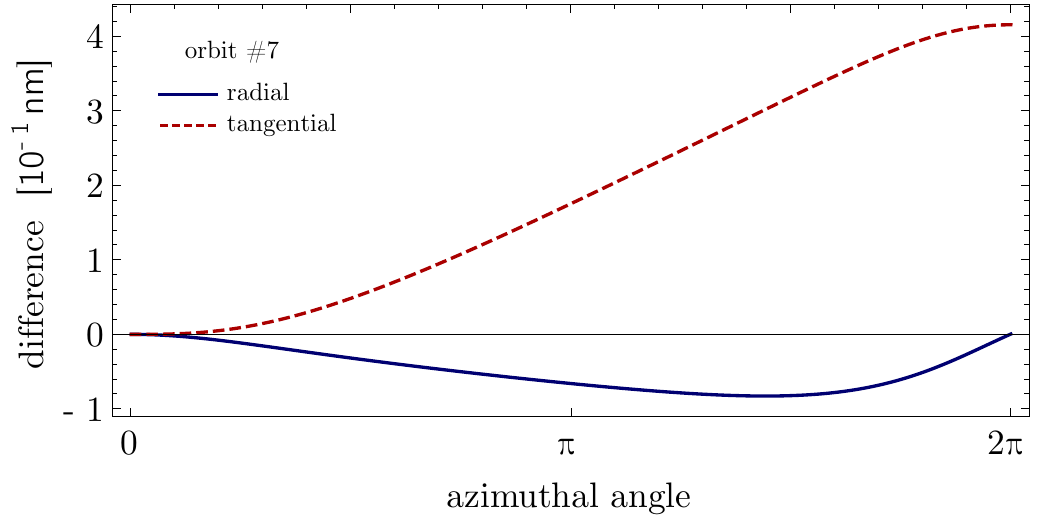}
	\caption{\label{Fig_HPS_pN_diff} The difference between the pN orbits (calculated using the XHPS) and the Schwarzschild orbits (calculated as geodesics of the Schwarzschild spacetime in isotropic coordinates and parametrized by coordinate time). The results are shown for orbit \#2 (smallest eccentricity, top), orbit \#6 (largest eccentricity, middle), and the LEO \#7, which is used in Fig.\ \ref{Fig_pNacc} as well, (bottom). In either case, we show the radial and tangential difference between both solutions, calculated at the same grid points.}
\end{figure}

Fig.\ \ref{Fig_HPS_pN_diff} shows the accuracy of the first-order pN orbits for three exemplary cases. We show the radial and tangential deviation from the solution of the geodesic equation in the Schwarzschild spacetime, which we constructed numerically with sufficient accuracy. To be consistent, we have to use isotropic radial coordinates and a coordinate time parameterization in either case. The maximal deviation is found for the elliptical orbit $\#6$, where it is in the nanometer regime. Hence, the result is as expected and summarized by: the first-order pN approach yields orbits at the nm accuracy level for satellite constellations around the Earth.

The way we assure identical initial conditions, at the level of accuracy that is inherent for the first-order pN approximation, for orbits in the Schwarzschild spacetime and its pN analogue is as follows. 
The orbits are confined to the equatorial plane, start at the perigee, and the argument of the perigee is taken to be zero. Both, the Schwarzschild orbit and the pN orbit are parametrized by the respective coordinate time, and the isotropic radial coordinates $\lambda$ and $R$ are used, respectively. We read off the eccentricity and semi-major axis, for which the Schwarzschild area coordinate $r$ is used, from Tab.\ \ref{Tab_orbits}. Then, the perigee and apogee radii are calculated and transformed to the Schwarzschild isotropic coordinate $\lambda$. We also calculate the initial azimuthal velocity w.r.t.\ the coordinate time by
\begin{align}
	\overset{\circ}{\Phi}_0 = \dot{\varphi}_0 \, \dfrac{c^2 A(r_0)}{r_0^2} \, .	
\end{align}
Hence, the non-vanishing initial conditions for the orbit in the Schwarzschild spacetime are given by $\lambda_0$ and $\overset{\circ}{\varphi}_0$. For the pN orbits, we identify the angular coordinates with those of the Schwarzschild spacetime, and for the radial coordinates we use $\lambda = R$ at the given level of accuracy. We then use the same numerical values for $R_0 = \lambda_0$ and the initial azimuthal velocity for the pN orbit integration with the XHPS.

The identification of the pN spherical coordinates with the Schwarzschild isotropic coordinates introduces an a priori error. Nevertheless, this error is of the order ${\mathcal{O}\big(c^{-3}\big)}$ and below the accuracy of the first-order pN approximation. From a mathematical point of view, we simply use the same initial values for a differential equation and its approximation.

\section{\label{Sec_RelativisticEffects} Magnitude of relativistic corrections}

To judge the importance and magnitude of relativistic effects, we compare the acceleration (and therefore the force) caused by the pN corrections in the EOM \eqref{Eq_pN_EOM_Cartesian} to various non-gravitational perturbations. We use the XHPS and take into account the SRP \citep{List:15}, the effects of Earth's albedo, the atmospheric drag, and the thermal radiation pressure (TRP) \citep{Rievers:16} since these appear to be the dominant effects. 

Here, all used non-gravitational force models are based on a detailed finite element (FE) model of the satellite, considering the orientation of each element with respect to the disturbance source, material properties as well as shadowing conditions with respect to satellite attitude \citep{List:15}. Albedo and Earth infrared (IR) models are based on hourly CERES data with a spatial resolution given by a fixed grid, $1^\circ \times 1^\circ$ latitude by longitude \citep{Wielicki:1996}. TRP \citep{Rievers:2011} is computed with a transient temperature model for each element of the FE satellite model with absorbed radiation from Sun, albedo and IR. The resulting effects due to Earth's atmosphere are modeled with a basic atmospheric drag model with constant drag coefficient. The atmospheric density is computed by using the empirical JB2008 model \citep{Bowman:2008}.

\begin{figure*}[]
	\centering
	\includegraphics[width=0.87\textwidth]{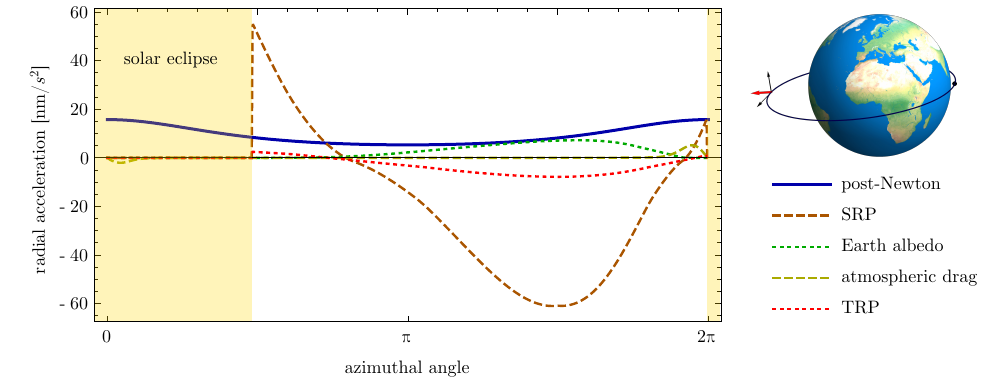}
	\includegraphics[width=0.87\textwidth]{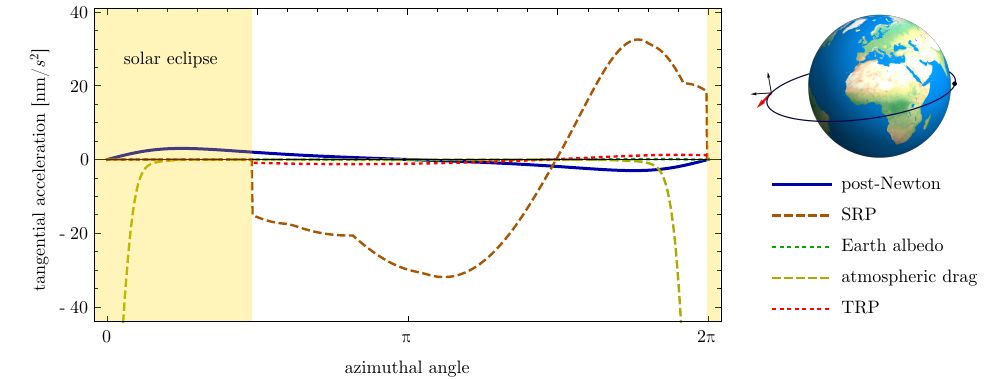}
	\includegraphics[width=0.87\textwidth]{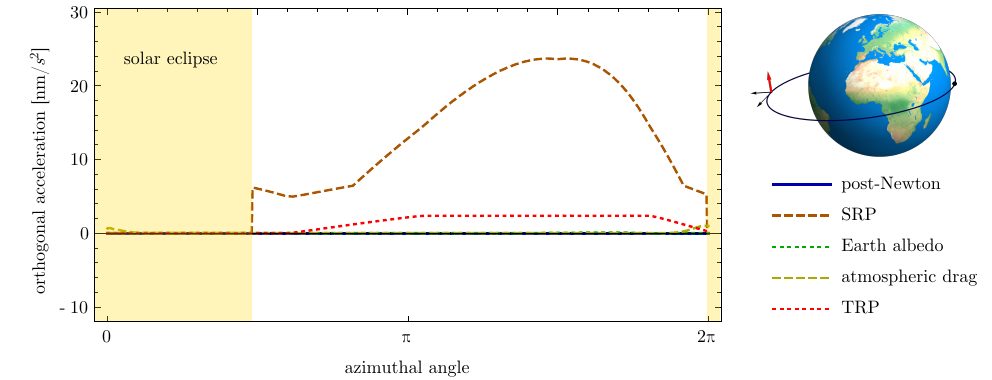}
	\caption{\label{Fig_pNacc} The magnitude of the accelerations due to the first-order pN contribution to the EOM, the solar radiation pressure (SRP), Earth's albedo, atmospheric drag, and the thermal radiation pressure (TRP). The orbit starts at the perigee and has an eccentricity $e \approx 0.2$ and a semi-major axis $a \approx 8.5\cdot 10^6\,$m. All accelerations were modeled using the XHPS, a GRACE-like model for the satellite properties, and Nadir pointing.}
\end{figure*}

\begin{figure*}[]
	\centering
	\includegraphics[width=0.87\textwidth]{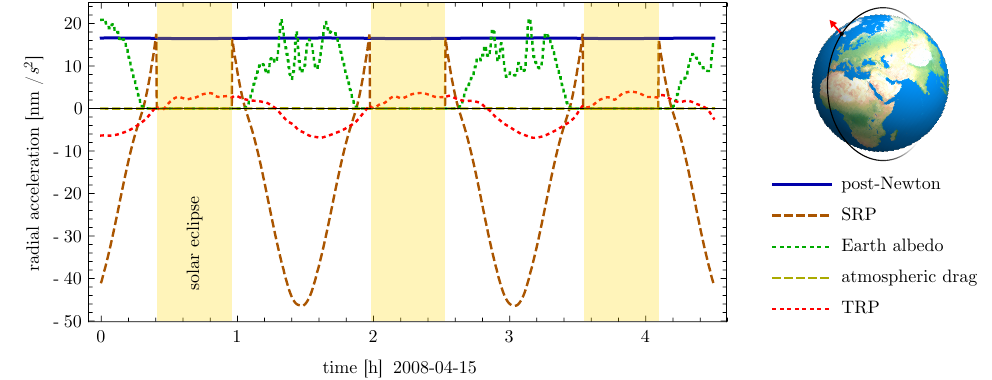}
		\caption{\label{Fig_pNacc_GRACE} The magnitude of the accelerations due to the first-order pN contribution to the EOM, the solar radiation pressure (SRP), Earth's albedo, atmospheric drag, and the thermal radiation pressure (TRP). We use a GRACE orbit and satellite attitude from 2008-04-15 and all accelerations were modeled using the XHPS with a GRACE FE model.}
\end{figure*}

We use a) the orbital test scenario $\#7$, see Tab.\ \ref{Tab_orbits}, and b) a real 24h GRACE orbit from 2008-04-15. For the satellite properties, we use a GRACE-like model with Nadir pointing. The result is shown for the case a) in Fig.\ \ref{Fig_pNacc} for the three perpendicular directions in a local Cartesian reference system along the satellite orbit: the radial direction, the tangential direction (in the momentary orbital plane) and the orthogonal direction. The accelerations due to the different effects are shown along the solution of the full EOM 
\begin{align}
	\mathbf{a} = \mathbf{a}_{\text{N}} + \mathbf{a}_{\text{pN}} + \mathbf{a}_{\text{env}} \, ,
\end{align}
where $\mathbf{a} = \overset{\circ \circ}{\mathbf{x}}$ is the acceleration vector in the GCRS. The Newtonian and pN accelerations ($\mathbf{a}_{\text{N}}$, $\mathbf{a}_{\text{pN}}$) are given by Eq.\ \eqref{Eq_pN_EOM_Cartesian_general}, and $\mathbf{a}_{\text{env}}$ contains the effects due to all non-gravitational (environmental) perturbations mentioned before. For the GRACE orbit (case b) we show the result in Fig.\ \ref{Fig_pNacc_GRACE}. Here, we only focus on the effects in the radial direction because relativistic effects are expected to have the largest contribution there.

We find the relativistic accelerations in the radial direction to reach a maximum of about $20\,$nm/s$^2$ in either case and they are, thus, comparable to the non-gravitational perturbations shown in the same figures. Hence, at least the first-order pN effects need to be taken into account to accurately model satellite orbits for high precision space missions \footnote{As high precision space mission we understand scenarios where orbital effects due to the space environment (SRP, Albedo, etc.) need to be taken into account to meet the accuracy goals of the respective mission.} in an environment around the Earth. The results shown here do agree with those shown in Tab.\ 3 of \citep{Soffel:2016}. In this work, the authors estimated the magnitude of several orbital effects, including pN contributions and the space environment, for the LAGEOS satellite.

Concerning orbit determination, we note the following. As can be seen in Fig.\ \ref{Fig_pNacc_GRACE}, the first-order relativistic contribution to the acceleration is almost constant along the circular GRACE orbit. Furthermore, the radial direction is most sensitive to relativistic effects. For orbit determination, in the POD process the effects could then easily be absorbed into other parameters if they are not correctly implemented in the orbit propagation model. However, it is not possible to make a statement on the magnitude of the residuals with and without pN corrections terms in the EOM at this point.

\section{\label{Sec_Conclusions} Conclusion}
The purpose of this work was to quantify the accuracy of different methods to approximately solve the general relativistic EOM. We have investigated different methods in the spherically symmetric Schwarzschild spacetime and its first-order pN approximation. Moreover, we have compared the magnitude of relativistic corrections in a pN spacetime to various non-gravitational perturbations of satellite orbits.

To solve the relativistic EOM in the Schwarzschild spacetime, we have used direct numerical integration, a semi-analytical Lie-series approach, and the exact analytical solution in terms of the Weierstrass elliptic function. The latter served as the reference solutions and enabled us to test the accuracy of the other methods. To obtain satellite orbits in the pN approximation of the Schwarzschild spacetime, we included relativistic corrections in the XHPS. 

When compared to the exact solution of the geodesic equation in the Schwarzschild spacetime, both, the direct numerical integration using a Runge-Kutta scheme on a fixed numerical grid, and the semi-analytical Lie-series approach yield (sub-)nanometer accuracies for a pre-defined set of test orbits. The drawback of using the Lie-series method is the longer computation time due to nested Poisson-brackets that are calculated analytically. However, the analytical part of the Lie-series approach may yield further insight into the dynamics in the spectral domain. Hence, this method may turn out to be an important link between the numerical integration and analytical solutions in later studies.

The first-order post-Newtonian approximation of the Schwarzschild spacetime was considered and the equations of motion are found to be modified Keplerian orbital equations. We solved these equations by implementing the relativistic corrections into the XHPS. The results verified the accuracy of the post-Newtonian approximation to the nanometer level, and we have shown that for a GRACE-like satellite in a low Earth orbit, the relativistic acceleration is comparable to various environmental perturbations. Hence, relativistic effects need to be taken into account for high precision space missions.

In a follow-up paper, we analyze longer orbital arcs and the relativistic orbital effects in a pN approximation of a more complicated gravitational field of the Earth. We will consider higher multipole moments in the Newtonian gravitational potential and analyze GRACE, GRACE-FO, TOPEX, LAGEOS, and other satellite constellations to compare the relativistic effects to gravitational (other bodies) and non-gravitational (environmental) perturbations. Since the focus of the present paper was purely orbit modeling, we will also take into account the relation to orbit determination in future work.
Furthermore, we will investigate extended satellites and the coupling of their moments of inertia to the gravitational field with its higher order multipoles. For such a situation, there is no analytical solution to the general relativistic problem. However, the presented in this work for the spherically symmetric case give strong confidence in the applicability and accuracy of the pN approximation of GR and the use of semi-analytical and numerical integration tools for orbit propagation, such as the XHPS, to solve the post-Newtonian equations of motion. We will also analyze how to determine the relativistic geoid \citep{Philipp:2017a} and properties of the relativistic gravitational field by GRACE-like mission scenarios.


\section*{Acknowledgement}
The present work was supported by the Deutsche Forschungsgemeinschaft (DFG) through the Sonderforschungsbereich (SFB) 1128 Relativistic Geodesy and Gravimetry with Quantum Sensors (geo-Q) and the Research Training Group 1620 Models of Gravity. We also acknowledge support by the German Space Agency DLR with funds provided by the Federal Ministry of Economics and Technology (BMWi) under grant number DLR 50WM1547.

The authors gratefully acknowledge insightful discussions with J\"urgen M\"uller, Volker Perlick, and Dirk P\"utzfeld.

\section*{References}

\end{document}